\documentclass[pdflatex,sn-mathphys-ay]{sn-jnl}

\usepackage{graphicx}%
\usepackage{epstopdf}
\usepackage{multirow}%
\usepackage{amsmath,amssymb,amsfonts}%
\usepackage{amsthm}%
\usepackage{mathrsfs}%
\usepackage[title]{appendix}%
\usepackage{xcolor}%
\usepackage{textcomp}%
\usepackage{manyfoot}%
\usepackage{booktabs}%
\usepackage{listings}%
\usepackage{fancyhdr}

\begin{document}
\title[An Autonomous Approach to Model Daytime Behavior of Sub-Ionospheric VLF Signals over Short and Medium Propagation Paths in India]{An Autonomous Approach to Model Daytime Behavior of Sub-Ionospheric VLF Signals over Short and Medium Propagation Paths in India}


\author[1]{\fnm{Sayak} \sur{Chakraborty}}\email{sayak.kolkata@gmail.com}

\author*[1]{\fnm{Tamal} \sur{Basak}}\email{tamalbasak@gmail.com}

\author[1]{\fnm{Sourav} \sur{Palit}}\email{souravspace@gmail.com}

\author[1]{\fnm{Sandip} \spfx{K.} \sur{Chakrabarti}}\email{sandipchakrabarti9@gmail.com}

\affil[1]{\orgdiv{} \orgname{Indian Centre for Space Physics}, \orgaddress{\street{466 Barakhola, Netai Nagar}, \city{Kolkata}, \postcode{700099}, \country{India}}}

\abstract{We present an autonomous model to simulate the daytime variation of sub-ionospheric Very Low Frequency (VLF) signal amplitude, beginning with the computation of the D-region electron density by numerically solving the electron continuity equation (ECE). From the resulting altitude-dependent electron density profile ($N_e$), we extract Wait's ionospheric parameters ($h^{\prime}$ and $\beta$) using a log-linear fitting method. The study focuses on two VLF propagation paths (one short and one medium in length) in India, originating from the VTX / 18.2 kHz transmitter. The model effectively employs the Long Wave Propagation Capability (LWPC) framework to reproduce the daytime VLF signal amplitude profile. It accurately captures the daytime variations observed at the Bengaluru (BAN) station, where the ground wave component is dominant, as well as at Khukurdaha, WB (KHK). A quantitative comparison between the simulated ($A_{sim}$) and ($A_{obs}$) observed amplitudes shows justified agreement, validating the physical consistency and possible predictive capability of the proposed approach.}

\keywords{Daytime variation in VLF signal, D-region ionosphere, numerical modeling of ionosphere, solar EUV and X-ray modeling}

\maketitle
\thispagestyle{fancy}
\section{Introduction}\label{sec1}

Sub-ionospheric Very Low Frequency (VLF) radio signal propagation is one of the most widely used tools for studying the D-region ionosphere. The primary energetic source responsible for ionizing the D-region is solar radiation. This region consists mainly of molecular oxygen, molecular nitrogen, and other compounds with varying ionization cross sections. Nitric oxide, which has an ionization potential of approximately 9.25 eV \citep{chakraborty24}, is a notable constituent of the upper part of the D-region ionosphere. Lyman-$\alpha$ radiation from the Sun is primarily responsible for ionizing the upper part of this region. However, solar Extreme Ultraviolet (EUV) radiation (wavelength $\sim 5$–$130$ nm) \citep{nicolet60,torr79,torr85}, along with X-rays, can penetrate deeper and ionize the D-region. Hence, both EUV and X-rays contribute significantly to D-region ionization \citep{chakraborty24}. During solar quiet periods, EUV becomes a more dominant source of ionization. The sub-ionospheric VLF signal carries valuable information about the D-region ionosphere as it propagates from the transmitter to receiver through the Earth-ionosphere waveguide (EIWG). These propagation paths exhibit distinct characteristics depending on their geographical locations, geodesics ($\rho$), and several other parameters \citep{thomson01, zigman07, pal10, schmitter11, chakraba12b, basak13, palit13, nina18}. The Long Wave Propagation Capability (LWPC) \citep{Ferguson1998, ferguson98} code is a widely used tool for numerically simulating sub-ionospheric VLF signals based on these parameters by solving modal equations. It is a freely available modeling tool, originally developed by the U.S. Naval Ocean Systems Center (NOSC), and has since been extensively applied in ionospheric and VLF propagation research. Many research groups have extensively employed the LWPC model to investigate sub-ionospheric VLF signal propagation under varying ionospheric conditions. This numerical tool has proven to be a reliable framework for modeling the waveguide characteristics of the Earth–ionosphere system and for interpreting perturbations associated with solar and geophysical events. For example, \cite{Pal12} modelled the effects of the Total Solar Eclipse of 2009 over multiple paths within India using LWPC. They simulated observed amplitude enhancements and reductions, and estimated rise in lower‐ionospheric reflection height ($h^\prime$) for VTX–Malda and VTX–Kolkata path). \cite{palit13} combined GEANT4 Monte Carlo‐derived ionization profiles with LWPC to reproduce VLF signal‐profile perturbations during M‐class and X‐class solar flares along the NWC - IERC/ICSP (India) path. \cite{Chand23} analyzed the signatures of early VLF events on the NWC-Suva (Fiji) path and identified the causative lightning via WWLLN data. They distinguished early/fast and early/slow onset events, and modeled amplitude and phase perturbations using LWPC v2.1 to retrieve D-region Wait parameters ($h^\prime$, $\beta$).

Solar radiation directly impacts the D-region, and VLF signals vary according to the D-region ionospheric variations. As a result, significant daytime variation is observed in the VLF signal amplitude (and phase) profiles. \cite{chakrabarti12} reported observations of daytime variability for different signal propagation paths across the Indian subcontinent for the VTX /18.2 kHz transmitter signal. We develop an autonomous model to simulate VLF signals from the Indian naval VLF transmitter VTX / 18.2 kHz, located at Vijayanarayanam, Tamil Nadu, India ($8^\circ 23^\prime$ N, $77^\circ 45^\prime$ E). We apply this model to simulate the signals received at two different stations over the Indian subcontinent, namely: (i) Bengaluru (BAN, $12^\circ 58^\prime$ N, $77^\circ 38^\prime$ E), representing a short path, and (ii) Khukurdaha, WB (KHK, $22^\circ 27^\prime$ N, $87^\circ 45^\prime$ E), representing a medium-length path, on 20 December 2008, and compare the results with observed data. We find reasonable agreement. Interestingly, our simulated VLF signal for BAN exhibits good agreement with observations, despite its proximity to the VLF signal transmission skip zone from VTX transmitter.

In contrast to earlier approaches that rely on empirical models such as the International Reference Ionosphere (IRI) \citep{chowdhury21a}, assume fixed ionospheric reflection parameters ($h^{\prime}$ and $\beta$) for VLF simulations \citep{chakrabarti12}, or employ chemical models to estimate D-region electron density profiles \citep{palit13}, the present work develops a fully physics-informed, first-principles framework. Specifically, electron densities are derived by numerically solving the electron continuity equation (ECE), explicitly incorporating ion-chemical processes, solar EUV (including Lyman-$\alpha$) and X-ray ionization, and recombination dynamics. From these self-consistent profiles, Wait’s model parameters \citep{wait64} are extracted dynamically and coupled with the Long Wave Propagation Capability (LWPC) code to reproduce VLF amplitude variations. By linking the governing ionospheric chemistry with radio wave propagation autonomously in an unified scheme, the model advances beyond existing semi-empirical and chemical approaches, offering a more realistic and temporally resolved description of D-region control on sub-ionospheric VLF propagation. This integrated, path-dependent treatment represents a novel contribution to recent VLF propagation studies.

\section{Modeling and Data}\label{sec2}

\begin{figure}[h]
  \centering
  \includegraphics[scale=0.3,keepaspectratio]{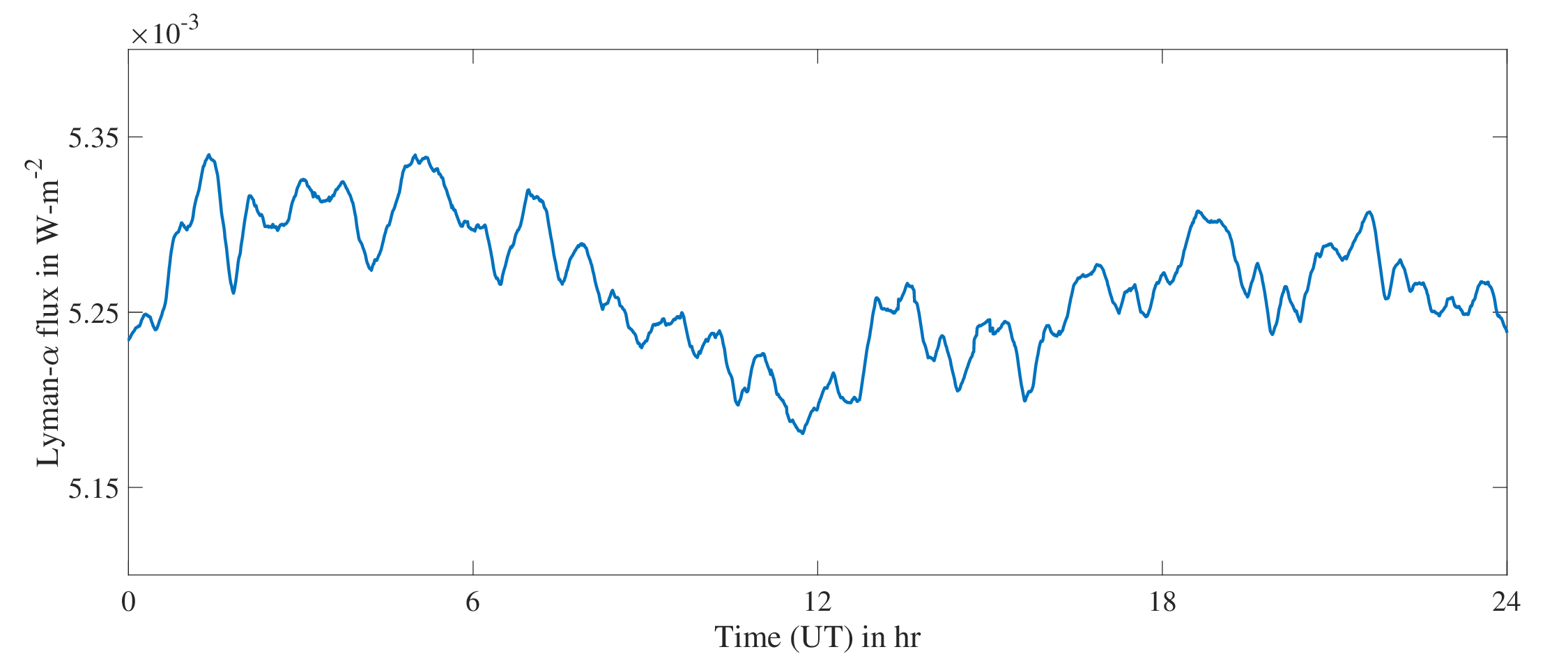}
  \caption{24 hours Temporal ($t$) profile of the solar Lyman-$\alpha$ flux on 20 December 2008.}
  \label{lyman}
\end{figure}

The D-region electron continuity equation (ECE) \citep[etc.]{whitten61,zigman07,basak13,palit15,nina18,chakraborty20} is as follows, 
 \begin{eqnarray}  \label{contunuity_eqn}
\frac{dN_e(t,h) }{dt} = \frac{q(t,h)}{1 + \lambda(h)} - \alpha_{eff}(t,h) N_e^2(h,t),
 \end{eqnarray}
where $\alpha_{eff}(t,h)$ represents the effective recombination coefficient, $\lambda(h)$ is the negative ion to $N_e(t,h)$ ratio and $q(t,h)$ is the total rate of ionization for solar EUV flux ($\phi_{euv}$) and soft X-ray flux  ($\phi_{xr}$). We start by solving ECE with appropriate dynamic inputs, such as solar X-ray and EUV, to obtain the electron density profile of the D-region. We compute the rate of ionization ($q(t,h,\rho)$), a significant term in ECE, at any given time $t$ and altitude $h$ for these different types of solar energy fluxes using Chapman’s formula \citep{chapman31}. Specifically, we divide each of the propagation paths into small segments ($\sim200$ km) and compute $q(t,h,\rho)$ over each of those segments. \citet{hayes17} estimated $\alpha_{\mathrm{eff}}(h)$ across the D-region during nine C-class solar flares along the NAA–Birr (Ireland) VLF propagation path. Following a similar methodology, we compute $\alpha_{\mathrm{eff}}(t,h)$ using altitude-dependent profiles reported by \citet{hayes17} and \citet{chakraborty20}, but adapted to the propagation paths considered in this study. The corresponding $\lambda(h)$ profiles are taken from \citet{palit15} and \citet{chakraborty22c}. The D-region electron density ($N_e(t,h,\rho)$) is then computed by solving ECE (equation ~\ref{contunuity_eqn}). A similar approach was followed by \cite{chakraborty24} to estimate $N_e(t,h)$, incorporating the contributions of both $\phi_{euv}$ and $\phi_{xr}$ as ionizing agents. We take the 15-second averaged solar light curve for photons within the 26–34 nm wavelength range, as obtained from the Solar EUV Monitor (SEM) onboard the Solar and Heliospheric Observatory (SOHO). We use the standard EUV spectrum ($\sim5$–$130$ nm) from \cite{torr79} and \cite{torr85} to calculate the solar EUV flux ($\phi_{euv}$). This spectrum consists of photon counts not only in the continuous range but also at discrete values corresponding to various lines, including the solar Lyman-$\alpha$ line (121.56 nm). The Lyman-$\alpha$ flux, for the entire day of our observation is plotted in figure ~\ref{lyman}. Additionally, the solar X-ray flux ($\phi_{xr}$) used in this study corresponds to a constant background level obtained from GOES observations, as the day under investigation exhibited minimal solar variability. The flux remained nearly constant at approximately $5.5 \times 10^{-9}~\mathrm{W\,m^{-2}}$.
\begin{figure}[h]
  \centering
  \includegraphics[scale=0.4,keepaspectratio]{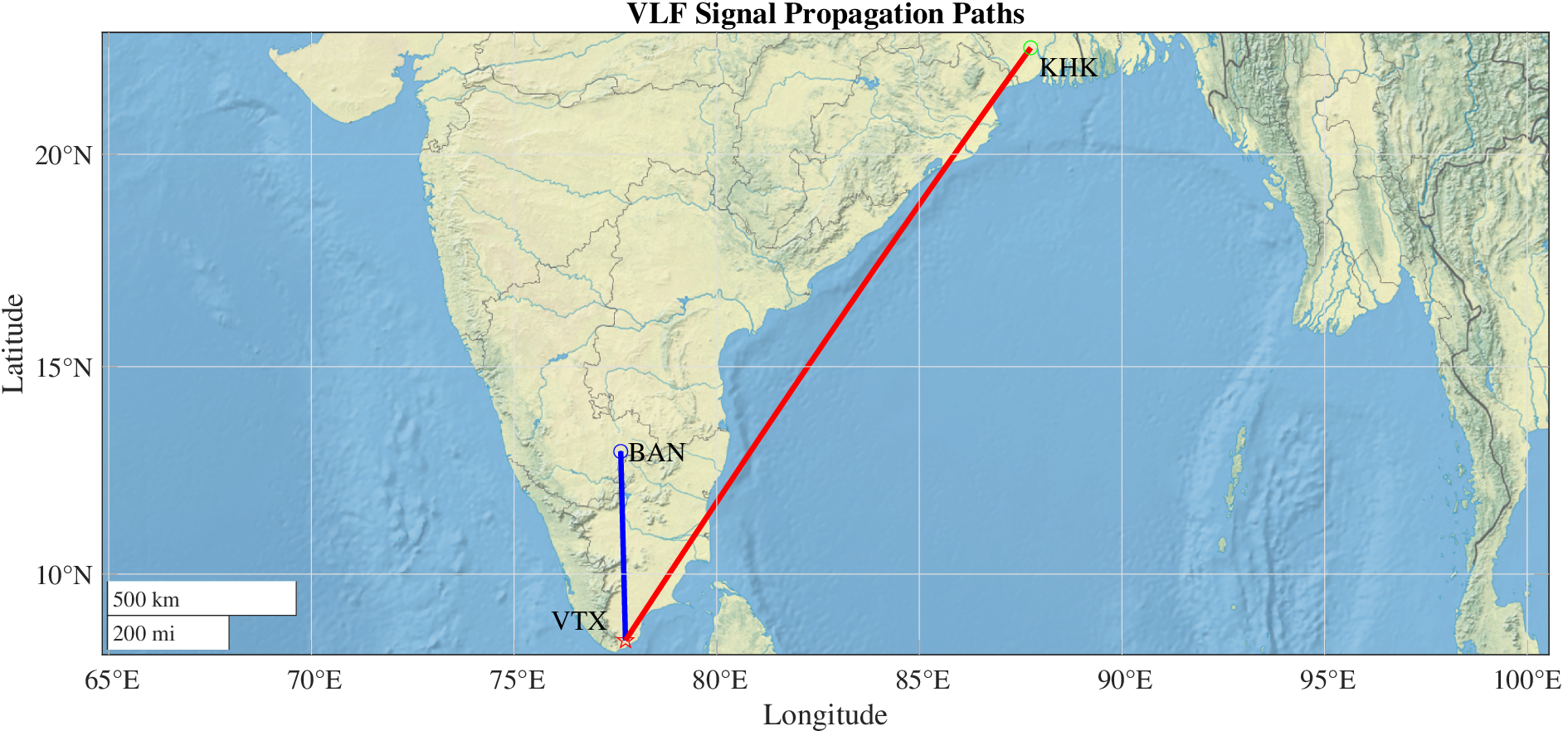}
  \caption{VLF signal propagation paths (i) VTX-BAN (blue) and (ii) VTX-KHK (red) on the map.}
  \label{map}
\end{figure}
Next, we use the log-linear extraction method \citep{chowdhury21a} to determine Wait’s ionospheric reflection parameters—$h^{\prime}(t,\rho)$ and $\beta(t,\rho)$—over each segment of the propagation paths. In general the Wait's parameters $h^{\prime}$ and $\beta$ represent effective reflection height and sharpness factors of the D-region ionosphere. For each time instant ($t$), we perform a log-linear fitting of the altitude profile of $N_e(t,h)$ \citep{chowdhury21a}. When $N_e$ is plotted as a function of $h$ at a fixed time, the variation exhibits a nearly linear trend in logarithmic scale. A linear fit exhibits the value of $\chi^2$ remains within the range between $0.9$-$0.98$ mostly. From this fit, the slope provides the value of $\beta$, while the intercept corresponds to $h'$. To formalize this, we employ Wait's empirical formula \citep{wait64}, expressed as  
\begin{eqnarray}  \label{wait}
N_e = 1.43\times10^{13} \exp(-0.15h^{\prime}) \exp[(\beta-0.15)(h-h^{\prime})],
\end{eqnarray}
where $N_e$, $h^{\prime}$, and $\beta$ are expressed in SI units. This procedure is systematically repeated to obtain the temporal evolution of $h'$ and $\beta$ throughout the daytime, which are subsequently used as inputs to the LWPC model. 

Finally, we simulate the VLF amplitudes ($A_{sim}$) using the $h^{\prime}(t,\rho)$ and $\beta(t,\rho)$ profiles in the RANGE exponential sub-program of LWPCv2.1 framework which solves the modal equations. By repeating the entire aforementioned procedure at suitable time intervals, we are able to autonomously generate the VLF signal amplitude profile from sunrise to sunset. During the simulation, we provide approximate profiles of the relevant input parameters, such as the solar zenith angle ($\chi$), effective recombination coefficient ($\alpha_{eff}$), mesospheric temperature ($T$), conductivity of the Earth’s surface ($s$), D-region effective collision frequency ($\nu$), etc., as available in the literature for similar signal propagation paths and comparable geophysical conditions. At present, our model is capable of simulating only the daytime VLF signal profile; therefore, we refrain from modeling the nighttime VLF signal amplitude.

The combination of (i) ECE solutions, (ii) extraction of Wait's parameters from $N_e$ and (iii) use of LWPC framework for generating VLF signal is a self-complete process for simulating sub-ionospheric VLF signals—starting from the solar energetic inputs to the D-region, hence, we refer to it as an autonomous model.

The VLF signal amplitude data were recorded using magnetic loop antenna and portable receiver systems during the winter VLF campaign across Indian sub-continent organized by the Indian Centre for Space Physics (ICSP), Kolkata during December 2008 \citep{chakrabarti12}. Then, the VLF signal from the VTX / 18.2 kHz was monitored across dozens of stations simultaneously.  We consider the VLF data recorded on 20 December 2008 at the two designated receiving stations BAN (Bengaluru) and KHK (Khukurdaha, WB) which were among the many stations used during the campaign. On that day, no solar flares, solar particle events, geomagnetic storms, or seismic activity were reported. Consequently, an undisturbed daytime variation of the VLF signal was observed along both propagation paths. Finally, we compare the VLF signal amplitudes simulated for those stations with the observed data.

\section{Result and Discussions}\label{sec3}

\begin{figure}[h]
  \centering
  \includegraphics[scale=0.3,keepaspectratio]{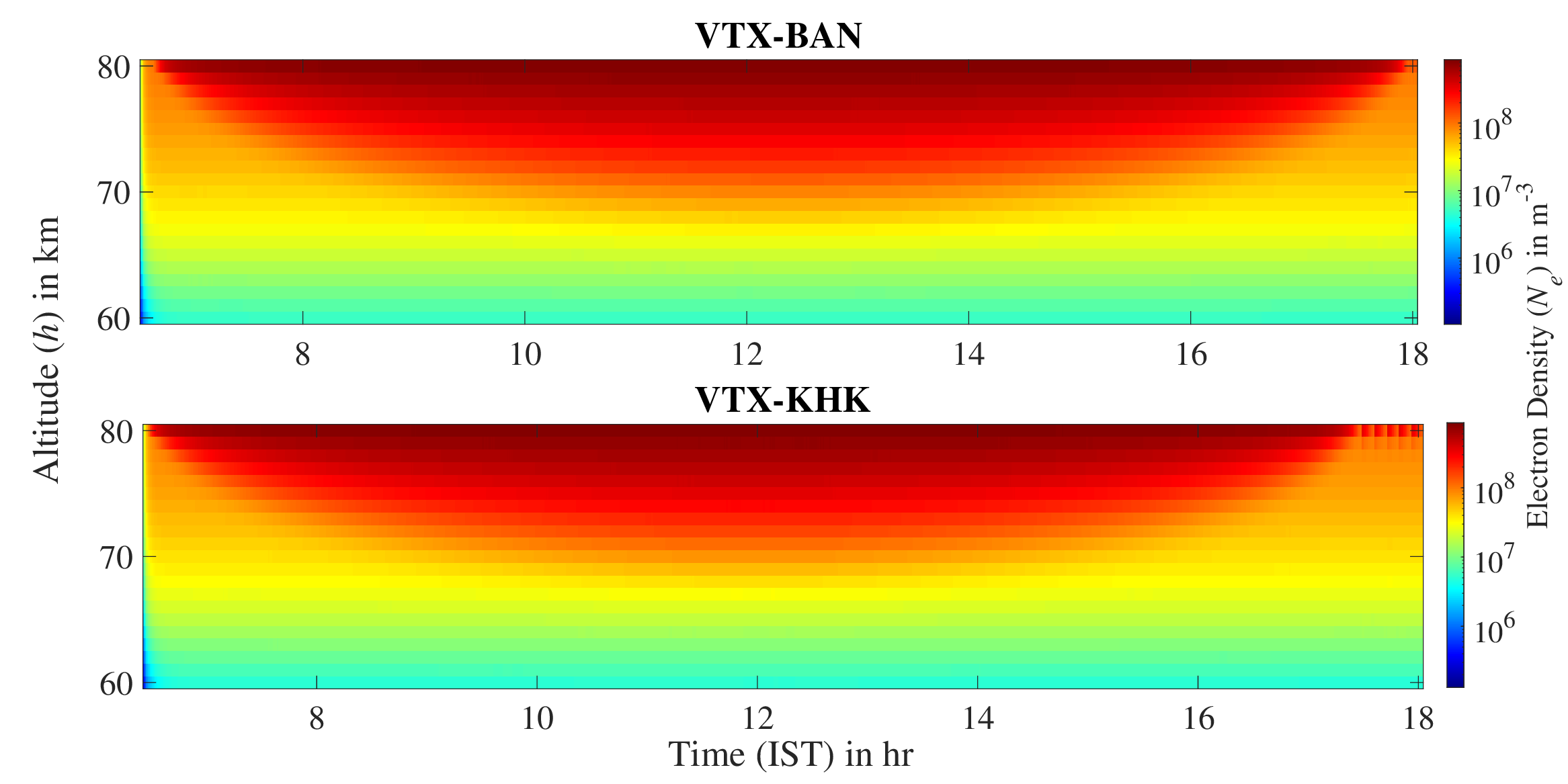}
  \caption{Altitude ($h$)-temporal ($t$) profiles of D-region electron density ($N_e(t,h)$) during daytime over the middle of the respective propagation paths VTX-BAN (top) and VTX-KHK (bottom).}
  \label{neth}
\end{figure}

Firstly, we computed the altitude–temporal profile of $N_e(t,h)$ using ECE over the two signal propagation paths. The values of $N_e(t,h)$ at different ionospheric altitudes are in close agreement with the standard values estimated by the International Reference Ionosphere (IRI) or other similar observational techniques. The $N_e(t,h)$ profiles at the midpoints of those propagation paths are presented in Fig.~\ref{neth}. As soon as the Sun rises, the D-region begins to form by absorbing necessary solar radiation, and $N_e(t,h)$ starts increasing gradually approximately from $10^6$ m$^{-3}$. During local noon, the solar zenith angle $\chi(t)$ reaches its minimum value, resulting in a maximum $N_e(t,h)$ values within $10^8$ m$^{-3}$ and $10^9$ m$^{-3}$. Thereafter, it gradually decreases again to approximately $10^6$ m$^{-3}$ near sunset. Additionally, we note a systematic enhancement in $N_e(t,h)$ values with increasing altitude at all times. Such $N_e(t,h,\rho)$ profiles are similar for both propagation paths. However, just before sunset (17-18 Hrs. IST), an oscillation in $N_e(t,h)$ is observed for the VTX–KHK propagation path, which is naturally the longest one between the two paths. Close to $\chi(t) \sim 90^\circ$, the D-region ionosphere nearly vanishes above the major part of the propagation path. As a result, a mixture of daytime and nighttime conditions appears in that region. Under this circumstance, additional modal interference patterns generated at that time may be a possible reason for the observed oscillations in $N_e(t,h)$ (Fig.~\ref{neth}).

\begin{figure}[h]
  \centering
  \includegraphics[scale=0.3,keepaspectratio]{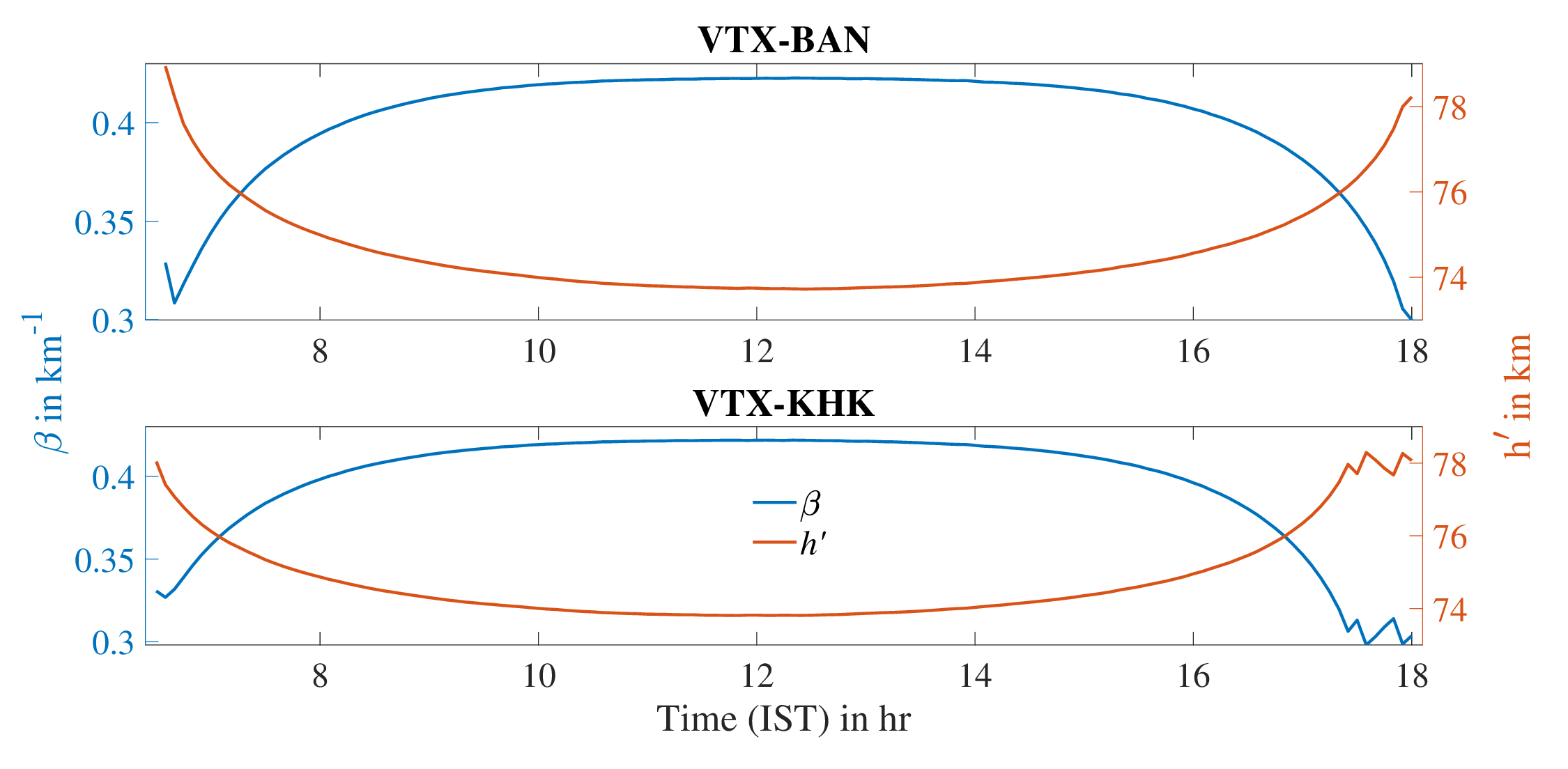}
  \caption{Temporal profiles of Wait's parameters $h^{\prime}(t)$ and $\beta(t)$ as extracted from the $N_e(t,h,\rho)$ profiles shown in Fig.~\ref{neth} during daytime over the middle of the respective propagation paths VTX-BAN (top) and VTX-KHK (bottom).}
  \label{hbeta}
\end{figure}

The profile of $\beta(t)$, as extracted using the log-linear extraction method, increases with increasing $N_e(t)$ over each of the segments of the propagation paths. However, the opposite trend is observed in the case of $h^{\prime}(t)$. The value of $\beta(t)$ varies from 0.3 to slightly more than 0.4 km$^{-1}$ from sunrise to local noon and returns to around 0.3 km$^{-1}$ during sunset. Similarly, $h^{\prime}(t)$ varies approximately from 78 km to 74 km between sunrise and local noon and then eventually returns to 78 km by sunset. The $h^{\prime}(t)$ and $\beta(t)$ profiles at the midpoints of the respective propagation paths are shown in Fig.~\ref{hbeta}. 

\begin{figure}[h]
  \centering
  \includegraphics[scale=0.3,keepaspectratio]{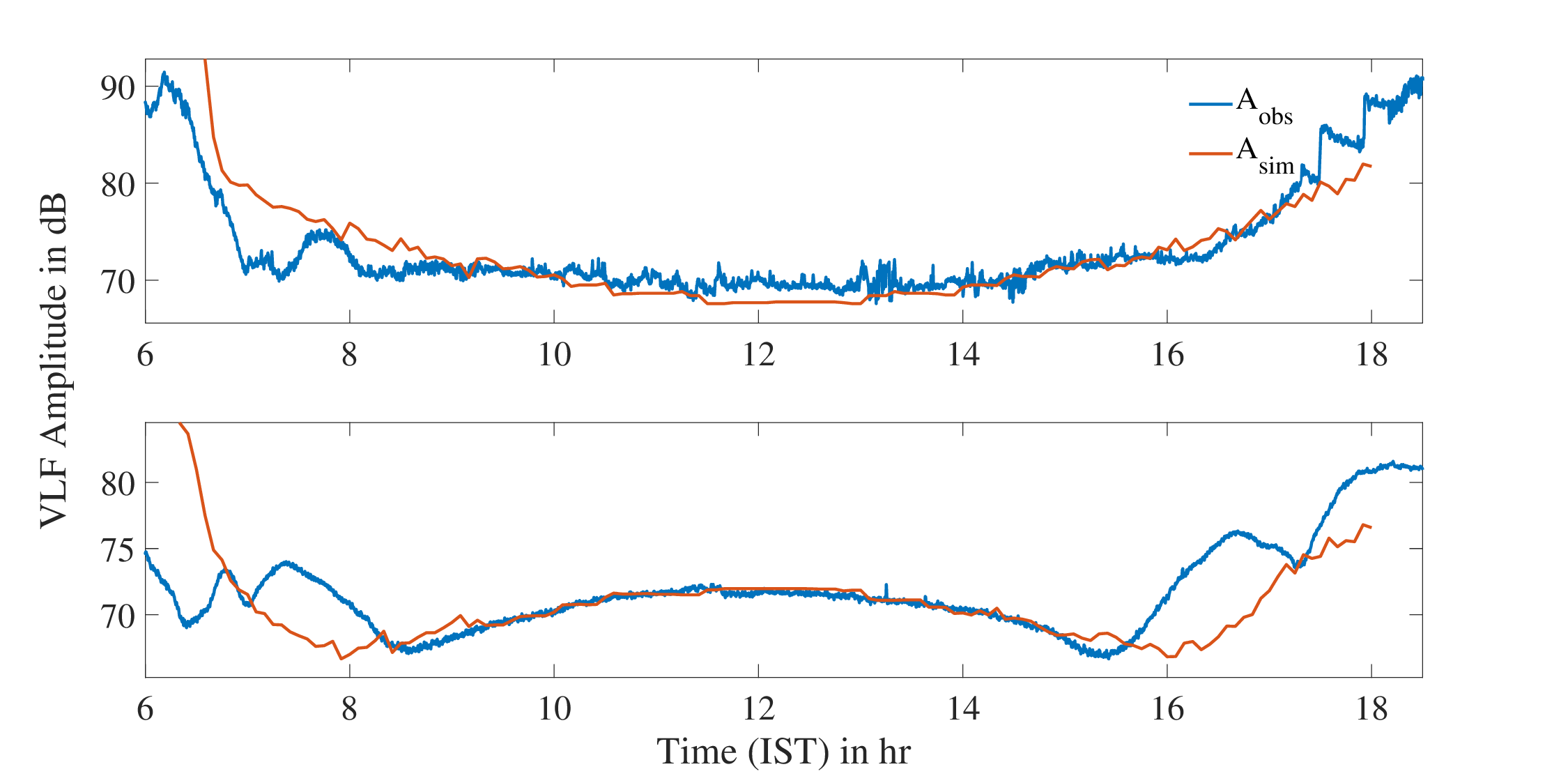}
  \caption{Daytime profiles of observed and simulated VLF signal amplitudes ($A_{obs}$ \& $A_{sim}$) for VTX-BAN (top) and VTX-KHK (bottom) propagation paths.}
  \label{vlf}
\end{figure}

We report the daytime variations of the simulated amplitude of VLF signal ($A_{sim}$), as obtained for BAN and KHK stations on 20 December 2008. The temporal profiles of $A_{sim}$ and $A_{obs}$ for the two chosen propagation paths are shown in Fig.~\ref{vlf}. We computed $A_{sim}$ in decibel (dB) units, relative to 1 $\mu$V·m$^{-1}$, which is similar to the units used in LWPC. Hence, we compare the numerical values of $A_{sim}$ and $A_{obs}$. Being in the vicinity to the skip zone of the VTX / 18.2 kHz transmitter, the daytime profile for the signal over VTX–BAN propagation path ($\sim$ 509 km) is generally formed due to interference of the dominant ground wave component of the signal. As a result, the $A_{obs}$ remains around 70 dB throughout the daytime. The typical daytime variation caused by mutual interference among ionospherically reflected sky waves is notably absent in $A_{obs}$. This characteristic is clearly evident in $A_{sim}$ (upper panel of Fig.~\ref{vlf}). For a major part of the daytime, $A_{sim}$ value is close to 70 dB. On the other hand, we note that $A_{sim}$ exhibits daytime variation in the case of VTX–KHK propagation path ($\sim$ 1894 km). Following $A_{obs}$, the $A_{sim}$ increases after sunrise and reaches to a maximum value during mid-day before coming back to a lower value during sunset. We performed the modal interferences among several sky-waves in the simulation process for the significant agreement between $A_{obs}$ and $A_{sim}$ (lower panel of Fig.~\ref{vlf}). 

In this study, specific nighttime lower ionospheric conditions are not incorporated into the model due to the absence of solar irradiance. Instead, a suitable constant $N_e(t,h)$ profile is employed. Consequently, we do not attempt to simulate the VLF signal amplitude profile during nighttime explicitly. Furthermore, we note that the agreement between $A_{sim}$ and $A_{obs}$ during the day–night terminator periods could be further improved.

The D-region is inherently complex to model due to its innumerable chemical components, highly variable ionization processes and sensitivity to multiple physical drivers. By employing an in situ autonomous framework, starting from solving the ECE and systematically incorporating ionospheric and geophysical parameters, we obtained a close agreement between $A_{\mathrm{sim}}$ and $A_{\mathrm{obs}}$. This agreement is not merely numerical but provides strong physical validation that the parameters used in the model more accurately represent the actual lower ionospheric conditions. Therefore, this model reestablishes sub-ionospheric VLF signal as a dependable tool for investigating remotely the D-region under solar energetic perturbations. Furthermore, the results offer valuable insight into the relative dominance of ground-wave and sky-wave components of the at different receiver locations for the same transmitter, thereby improving our understanding of signal propagation mechanisms in the D-region.

\section{Conclusion}\label{sec4}

In this study, we developed and implemented an autonomous model to investigate the daytime behavior of the D-region ionosphere and its effect on sub-ionospheric VLF signal propagation. Firstly, we solved ECE, accounting for ionization and recombination processes driven by solar EUV (including Lyman-$\alpha$) and X-ray background radiation. Secondly, from the resulting electron density profiles ($N_e$), Wait’s ionospheric parameters ($h^{\prime}$ and $\beta$) are extracted through a log-linear fitting approach. Lastly, we used the LWPC framework to simulate the amplitude variation of VLF signals ($A_{sim}$) across two Indian propagation paths: VTX–KHK (a medium-length path) and VTX–BAN (a short-length path). We intentionally selected two propagation paths with significantly different lengths to evaluate the model’s capability in handling varying ionospheric conditions and accurately reproducing VLF signal amplitudes under distinct modal scenarios. The model successfully reproduces the key features of the daytime VLF amplitude profiles ($A_{sim}$) and shows reasonable agreement with the observed data ($A_{obs}$) in both scenarios: (i) when ground wave propagation is dominant (VTX–BAN), and (ii) under conditions involving multiple interfering sky waves (VTX–KHK). While modest discrepancies are observed around the day–night terminator, they can be attributed to known limitations of the current approach, including the absence of post-sunset chemistry and ionospheric dynamics driven by terrestrial energy sources. Nevertheless, the model proves capable of self-consistently simulating multi-path VLF propagation using an electron density profile derived from ECE.

Looking ahead, we plan to extend this framework to include nighttime conditions by accounting for additional ionospheric processes, such as recombination and non-local photo-ionization, atmospheric gravity wave effects etc. We also aim to incorporate the influence of solar energetic events (e.g., solar flares and proton events), which will enable the simulation of disturbed conditions during solar active days. These future enhancements will broaden the model's applicability and improve its accuracy in capturing ionospheric variability across a range of geophysical environments.

At present, the model is capable of simulating daytime VLF signal propagation under quiet solar conditions for mid–low latitude paths, not limited to India. This ensures the framework already possesses global applicability under similar conditions, while the planned extensions will expand its usability further.

\backmatter

\bmhead{Acknowledgments}

The authors acknowledge the Indian Centre for Space Physics (ICSP), Kolkata, for providing access to the VLF signal data. Sayak Chakraborty gratefully acknowledges financial support from the DST-INSPIRE Fellowship (IF200266), Government of India. The authors also thank NCEI-NOAA for the solar X-ray data and the Laboratory for Atmospheric and Space Physics (LASP), University of Colorado, for the EUV light-curve data. Special thanks are extended to Mr. Debashis Bhowmick (ICSP) for his technical and scientific support related to the VLF data.

\bibliography{references}

\section*{Declarations}

\begin{itemize}
\item \textbf{Funding:} This work was supported by the DST-INSPIRE Fellowship (IF200266), Government of India.
\item \textbf{Conflict of interest/Competing interests:} The authors have no relevant financial or non-financial interests to disclose.
\item \textbf{Author contribution:}All authors contributed substantially to the conception and design of the study. Data collection was conducted by SC and TB. Data analysis and numerical simulations were performed by SC. The initial manuscript was drafted by SC, TB, and SP. The final version was prepared by SC, incorporating feedback and suggestions from all authors. The overall work was supervised by SKC, SP and TB. All authors reviewed and approved the final manuscript.
\item \textbf{Consent for publication:} All the authors give their concent on publication of this manuscript to the publisher.
\item \textbf{Data availability:} X-ray and EUV light-curve data are freely available from the NCEI-NOAA and the Laboratory for Atmospheric and Space Physics (LASP), University of Colorado, respectively. VLF data are available from ICSP upon reasonable request to the authority. The datasets, simulation outputs, and codes used in this study may be available to interested researchers upon reasonable request to the corresponding author. 
\item \textbf{Ethics declaration:} not applicable.

\end{itemize}

\end{document}